\documentclass[letterpaper]{article}

\usepackage[T1]{fontenc}

\usepackage{geometry}
\usepackage{setspace}

\usepackage[style = chem-acs]{biblatex}
\usepackage{graphicx}
\usepackage{float}
\usepackage{subcaption}
\newfloat{scheme}{htbp}{los}
\floatname{scheme}{Scheme}
\floatname{chart}{Chart}
\newfloat{graph}{htbp}{loh}

\usepackage{chemformula} % Formulas using \ch{}
\usepackage[version = 4]{mhchem} % Formulas using \ce{}

\usepackage{hyperref}
\usepackage{cleveref}
\usepackage{authblk}
\author[1]{Roee Asher}
\author[1]{Nadav Moav}
\author[1]{Lee A. Burton}
\affil[1]{Department of Materials Science and Engineering, The Iby and Aladar Fleischman Faculty of Engineering, Tel Aviv University, Ramat Aviv, Tel Aviv 6997801, Israel.}

\title{The Recurrent Structural Frameworks of Stable Inorganic Materials}
\date{*Email: leeburton@tau.ac.il}

\begin{document}

\maketitle

\begin{abstract}
The number of possible crystal structures vastly exceeds the number realized among thermodynamically stable inorganic materials, suggesting that experimentally accessible structure space is organized around a limited set of preferred structural frameworks. Analysis of 23,160 structures on the thermodynamic convex hull identifies 6,820 distinct structural frameworks, of which 2,382 recur across multiple chemically distinct materials and are therefore classified as structural prototypes. Within a filtered dataset, metallic systems exhibit particularly strong structural recurrence, with 6,270 materials grouped into 535 frameworks and 327 prototypes, while 5,557 ionic compounds occupy 1,704 frameworks and 696 prototypes. These results demonstrate that stable inorganic materials occupy a highly compressible region of structure space and establish a recurrence-based catalog of structural prototypes that provides an empirical measure of framework prevalence, enabling the relative likelihood of structural frameworks in future materials to be estimated from their recurrence among known compounds.

\end{abstract}

\section*{Keywords}

Atomic structure, crystallography, materials prototypes, high-throughput screening

%\section*{Abbreviations}

%Some journals require a list of abbreviations: these normally should be given immediately after the keyswords (if required).

\section{Introduction}

\begin{figure}[t!]
  \centering
  \begin{subfigure}[b]{0.32\textwidth}
    \centering
    \includegraphics[width=\textwidth]{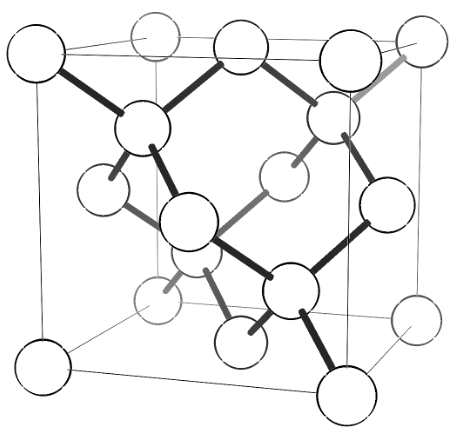}
    \caption{}
    \label{fig:diamond}
  \end{subfigure}
  \hfill
  \begin{subfigure}[b]{0.32\textwidth}
    \centering
    \includegraphics[width=\textwidth]{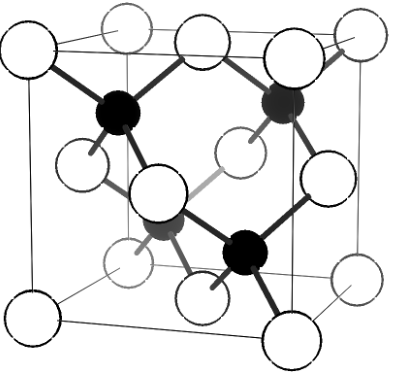}
    \caption{}
    \label{fig:zincblende}
  \end{subfigure}
  \hfill
  \begin{subfigure}[b]{0.32\textwidth}
    \centering
    \includegraphics[width=\textwidth]{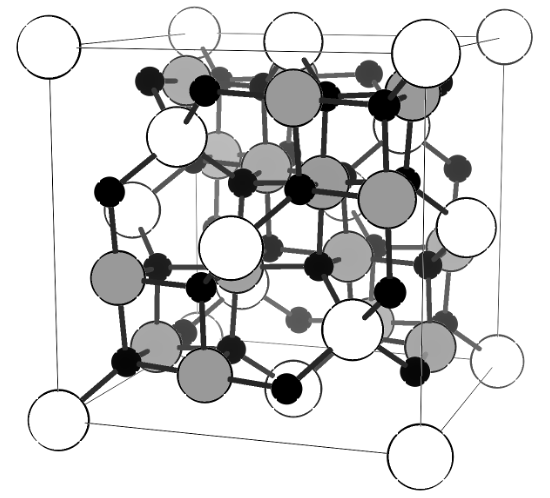}
    \caption{}
    \label{fig:spinel}
  \end{subfigure}

  \caption{Unit-cell structures of (a) diamond, (b) zinc-blende, and (c) spinel. These examples highlight both recurring geometric motifs and the challenges of conventional structure classification, motivating species-independent comparison of atomic arrangements. Different colors indicate distinct atomic species.}
  \label{fig:diamond-zincblende-spinel}
\end{figure}

Atomic structure fundamentally governs material properties~\cite{van_de_walle_complete_2008, cheng_atomic-level_2011, wang_influence_2024}. Materials with identical composition can exhibit markedly distinct behavior due to differences in atomic arrangement, known as allotropy in elements and polymorphism in compounds. Carbon for example, exists both as diamond, a tetrahedral structure with exceptional hardness~\cite{slack_thermal_1964, saslow_band_1966}, and as graphite, a layered structure with weak interlayer bonding that enables facile mechanical sliding~\cite{sanchez_egea_friction_2019, blakslee_elastic_1970}. Conversely, chemically distinct materials that adopt similar atomic structures often exhibit related physical and electronic characteristics. Diamond, \ce{ZnS}, \ce{GaAs} and \ce{InP} all crystallize in closely related tetrahedral frameworks (see \autoref{fig:diamond-zincblende-spinel}) with comparable mechanical and electronic behavior arising from the similar bonding environments~\cite{giesecke_1958_III-V_semiconductors, sze2007_physics}. These recurring geometric relationships suggest that many stable inorganic materials can be understood in terms of a relatively limited set of shared structures.

Traditional crystal structure classification schemes based on mineral analogies (\textit{e.g.}, referring to materials as having “zinc-blende” or “rock-salt” structures), spacegroups, and Pearson symbols are used to provide intuitive and standardized labels for shared structures. However, these descriptors become increasingly limited when extended across large datasets, where similar atomic arrangements can appear under different labels and distinct structures can share the same descriptor. Mineral-based names remain historically tied to particular compounds, which can be both undescriptive and confusing. For example, the rock-salt mineral is actually called halite, while the term zinc-blende may be used to describe materials not containing zinc. Symmetry-based descriptors such as space groups provide more rigorous crystallographic information, but they likewise do not uniquely describe structure. For example, both diamond and spinel crystallize in the $Fd\bar{3}m$ space group despite possessing different atomic structures (see again \autoref{fig:diamond-zincblende-spinel}). Pearson symbols encode only the crystal system, lattice centering, and the number of atoms in the unit cell. As such, structurally distinct arrangements may share identical Pearson symbols; for example, diamond and rock-salt possess the Pearson symbol \textit{cF8} despite exhibiting fundamentally distinct atomic arrangements and coordination environments. Each of these three descriptors summarizes particular crystallographic attributes rather than the atomic geometry, allowing structurally distinct materials to share the same descriptor while closely related structures receive different labels~\cite{putnis1992introduction, ewald_strukturbericht_1931, pearson_handbook_2013, villars_pearsons_1997, mehl2019brief, mehl2017aflow}.

A more general strategy for understanding inorganic structure space treats crystal structures as species-independent geometric frameworks that can be realized by multiple chemically compatible compositions. In this context, an atomic structure depends on the constituent species, whereas a structural framework refers to a geometric arrangement of sites independent of species.  The Inorganic Crystal Structure Database (ICSD) introduced a symmetry-based approach to structural classification in which frameworks recurring across multiple entries are designated as "structure types"~\cite{zagorac_recent_2019, allmann2007introduction}. However, recurrence within the ICSD does not necessarily correspond to independent chemical realizations, since multiple crystallographic refinements of the same material may be represented separately. Furthermore, structure-type assignment makes no distinction between thermodynamically stable compounds and high-energy polymorphs. Consequently, recurrence within the database does not necessarily reflect meaningful recurrence of low-energy structural frameworks in nature.

The Automatic-FLOW (AFLOW) database AFLOWlib incorporates descriptors such as stoichiometry, symmetry, and atomic coordinates to enable large-scale structural classification and organization of crystallographic data~\cite{curtarolo2012aflow, mehl2017aflow}. Such approaches provide broad structural coverage and facilitate comprehensive enumeration of known structure space. However, every geometrically unique framework is assigned as a distinct structure prototype (the equivalent of the ICSD structure types), regardless of whether it recurs across multiple materials. Treating every unique framework as a prototype makes prototype classification effectively indistinguishable from structure enumeration and becomes primarily an inventory of structural diversity rather than structural prevalence. Certain systems, such as hybrid organic--inorganic materials exhibit enormous configurational diversity arising from molecular degrees of freedom and framework flexibility. For example, metal--organic frameworks (MOFs) combine inorganic nodes with molecular linkers to generate vast numbers of distinct topologies~\cite{moosavi_understanding_2020, glasby_topological_2024}, while hybrid perovskites exhibit molecular orientation degrees of freedom that produce large numbers of structurally distinct realizations~\cite{Whalley2017}. In such systems, the distinction between a framework and a prototype ultimately disappears because prototype classification offers little compression of the underlying structure space. By contrast, stable inorganic solids are constructed from a comparatively limited set of atomic bonding motifs, suggesting that experimentally accessible inorganic structure space is highly compressible and therefore well suited to recurrence-based prototype classification.

\begin{figure}[t]
  \centering
  \hspace{-9mm}\includegraphics[width=\linewidth]{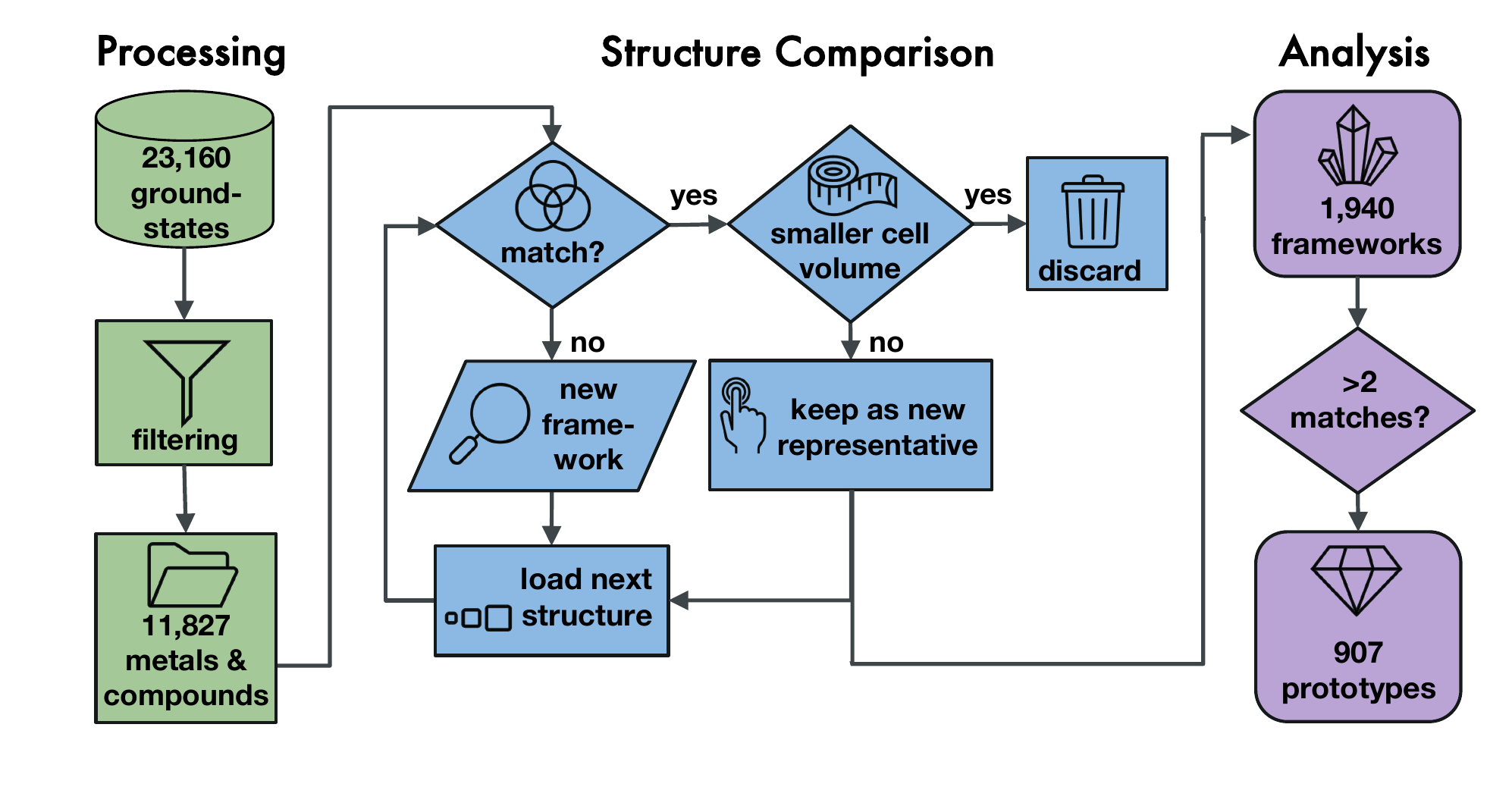}
  \caption{Workflow for identifying recurrent structural frameworks as prototypes. Crystal structures are compared using species-independent geometric matching and clustered into representative frameworks, from which recurring framework families are identified as prototypes.}
  \label{fig:figure2}
\end{figure}

Recent studies, including DeepMind's GNoME project, have predicted millions of new materials and tens of thousands of unreported structural prototypes~\cite{merchant_scaling_2023}. By emphasizing the exploration and enumeration of accessible structure space, such efforts also illustrate the limitations of defining every unique framework as a distinct prototype. From a recurrence-based perspective, structural novelty and prototype novelty are distinct: a newly identified structure may either represent a genuinely new prototype or a structural variant. The distinction is important because the utility of a classification scheme does not necessarily increase with the number of distinct structures it contains. The original Strukturbericht classification scheme sought to organize crystal structures through an expanding catalog of structure types, but the continued growth in the number of identified structures eventually led to the abandonment of new symbols~\cite{mehl2019brief}. Distinguishing structural novelty from prototype recurrence is therefore essential for understanding whether experimentally accessible inorganic structure space continues to expand through fundamentally new structural organizations or primarily through additional realizations of structural families;~\cite{cheetham_artificial_2024, leeman2024challenges} these families are the prototypes we attempt to identify. 
 
Prototypes in this way can enhance materials discovery. Candidate materials are commonly generated by mapping new compositions onto known structure types and subsequently evaluating their stability using methods such as density functional theory (DFT), a process known as element substitution~\cite{ceder_opportunities_2010, sun2016metastability, curtarolo_high-throughput_2013}. Likewise, crystal structure prediction methods such as AIRSS, USPEX, and XtalOpt exploit known structure types as seed configurations or reference motifs while exploring candidate atomic arrangements~\cite{pickard_ab_2011, oganov_crystal_2006, Oganov_2011_Evolutionary, XtalOpt}. Beyond identifying possible frameworks, prototype recurrence frequencies provide a quantitative prior for structure prediction, ranking candidate frameworks according to how frequently they are realized among stable inorganic materials. A prototype catalog could also provide compact representations of structure space for crystallographic organization, structural similarity benchmarking, and machine-learning representations of crystal structures. Such benchmarking studies are typically performed using \textit{ad hoc} prototype collections assembled for individual investigations. For example, Zimmermann and Jain benchmarked crystal structure similarity metrics using a curated set of 40 commonly studied crystal structures~\cite{zimmermann2020local}. The continued reliance on such \textit{ad hoc} collections, despite decades of structural classification efforts including Strukturbericht, the ICSD, and AFLOW, shows that a broadly adopted catalog of recurrent, thermodynamically stable framework families is still lacking.

%\begin{figure}[t]
  %\centering
  %\includegraphics[width=0.5\linewidth]{prototypes-paper/Figures/tolerances-example.png}
  %\caption{illustration of the three distinct tolerances assessed for the structure matching process.}
%  \label{fig:matching-nofilter}
%\end{figure}

In this work, we identify physically meaningful, recurrent structural frameworks arising from experimentally realized ground-state materials rather than enumerating all geometrically distinct structures. Using a curated set of 23,160 experimentally reported Materials Project structures predicted to lie on the thermodynamic convex hull~\cite{jain_commentary_2013, bartel2022review, sun2016metastability}, we employ the StructureMatcher algorithm implemented in the Python Materials Genomics (PyMatGen) toolkit~\cite{ong2013python} with a species-agnostic comparator to identify structural equivalence directly from atomic geometries rather than crystallographic metadata. The use of the Materials Project avoids duplicate experimental reports while allowing the selection of naturally occurring, thermodynamically stable frameworks. An efficient clustering procedure avoids exhaustive pairwise comparisons, enabling scalable analysis of large materials datasets. Applied to the full dataset, the workflow reduces 23,160 stable materials to 6,820 unique structural frameworks, of which 2,382 recur across multiple chemically distinct materials and are therefore classified as prototypes. The metallic subset contains 6,270 materials grouped into 535 frameworks and 327 prototypes, while the compound subset contains 5,557 materials grouped into 1,704 frameworks and 696 prototypes. Unlike existing approaches, which assign prototype status to every geometrically distinct structure, we define prototypes as structural frameworks that recur across multiple chemically distinct, thermodynamically stable materials. This shifts prototype classification from an inventory of structural diversity to a quantitative description of structural prevalence.

\begin{figure}[t]
  \centering
  \includegraphics[width=0.99\linewidth]{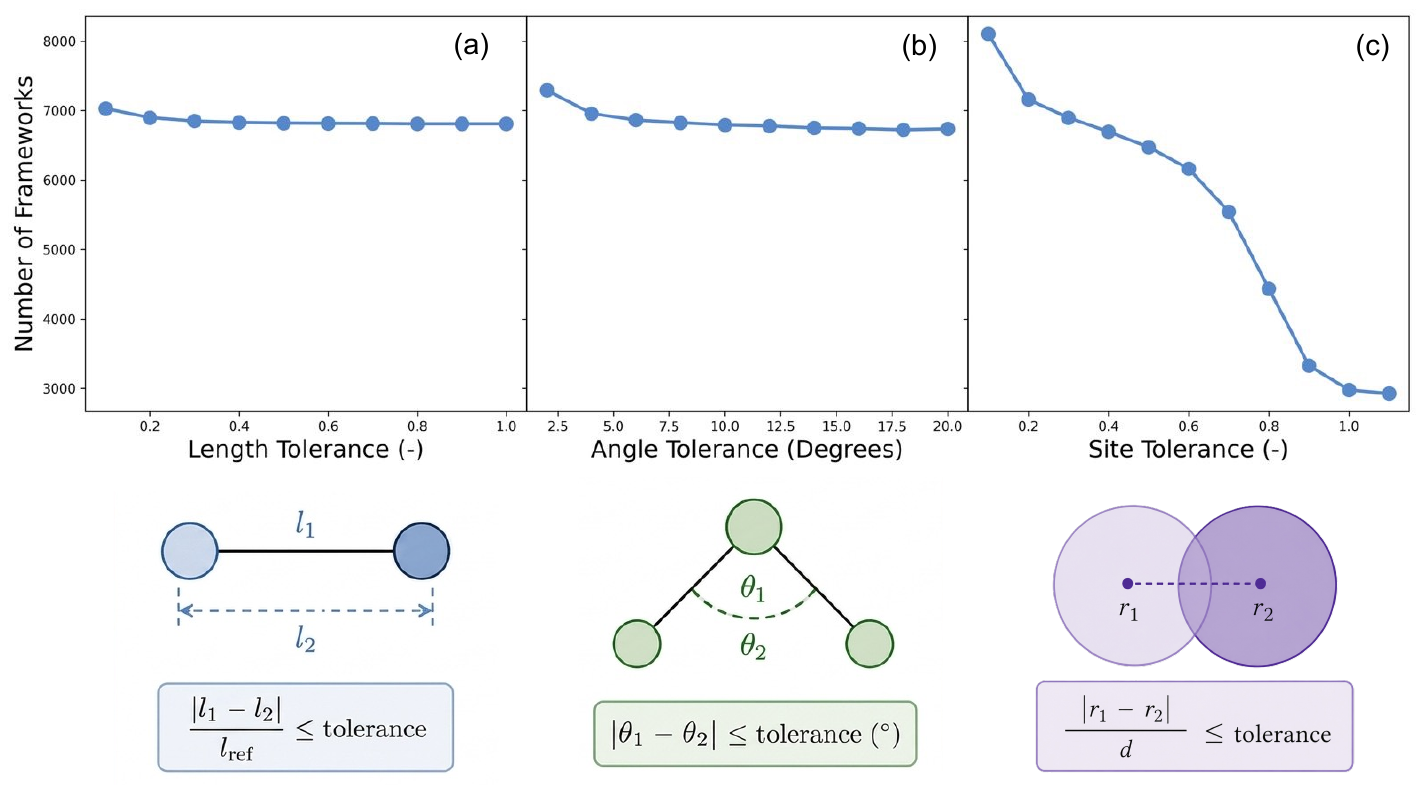}
  \caption{Sensitivity of framework identification to matching tolerances for the full dataset of 23,160 materials. The number of distinct frameworks identified is shown as a function of (a) length tolerance, (b) angle tolerance, and (c) site tolerance, according to the definitions underneath.}
  \label{fig:tolerance-sensitivity}
\end{figure}

\section{Methods}

All (and only) experimentally-observed, ground-state structures were obtained from the Materials Project database~\cite{jain_commentary_2013,bartel2022review, sun2016metastability}. Additional preprocessing excluded structures containing hydrogen, noble gases, structures with more than 100 atoms per unit cell, stoichiometric coefficients exceeding four, and structures containing more than three distinct cation or anion species. These restrictions define the intended scope of the study by focusing on chemically well-defined inorganic solids while maintaining computational tractability for large-scale structural comparison.

Structural equivalence was evaluated using the StructureMatcher algorithm implemented in PyMatGen~\cite{ong2013python}. Several crystallographic comparison tools were considered, including AFLOW-XtalFinder~\cite{hicks2021xtalfinder}, XTALCOMP~\cite{lonie2012identifying}, STRUCTURE TIDY~\cite{gelato1987structure}, CRYCOM~\cite{dzyabchenko1994method}, CMPZ~\cite{hundt2006cmpz}, SPAP~\cite{su2017construction}, and COMPSTRU~\cite{flor2016comparison, aroyo2011crystallography, aroyo2006bilbao1, aroyo2006bilbao2, tasci2012introduction}. StructureMatcher was selected for its open-source implementation and flexible control over geometric matching tolerances, making it well suited to the recurrence-based prototype identification workflow developed here. Species-independent matching was enabled through the FrameworkComparator, which compares occupied crystallographic sites while ignoring atomic identity. Two structures were considered equivalent when their atomic frameworks matched within specified tolerances for lattice parameters and atomic positions, independent of symmetry labels, space groups, or Wyckoff assignments~\cite{allmann2007introduction}. To avoid the quadratic scaling associated with exhaustive pairwise comparison, structures were compared iteratively against a dynamically updated set of representative frameworks, enabling scalable identification of prototypes across the complete dataset, as shown in \autoref{fig:figure2}. Representative structures used during framework matching were selected according to the largest unit-cell volume to minimize atomic overlap. Although alternative parameter choices or clustering procedures may lead to minor differences in framework membership, the overall picture of extensive structural recurrence appears robust. The framework and prototype counts reported here should therefore be interpreted as a reproducible estimate of structural organization rather than an exact or unique partitioning of inorganic structure space.

For categorization, chemical compositions were mapped onto species-independent stoichiometric formula types (ANX notation) following ICSD conventions~\cite{steudel_scientific_nodate}. This method groups materials according to stoichiometric pattern while remaining independent of elemental identity. Elements were assigned cationic or anionic roles according to their predominant oxidation behavior~\cite{ding2020}. Arsenic, which occurs in both cationic and anionic states, was consistently treated as a cation due to its dominant prevalence in the ICSD. With this information, the surviving candidates could be separated into compounds, metals/intermetallics, or miscellaneous systems. If a system exhibited a clear cation--anion partitioning it was classified as a compound; if only cationic species were present it was classified as metallic/intermetallic; and if only anionic species were present it was classified as miscellaneous. Thus, for example, \ce{CaAl4} is identified as a metallic/intermetallic system, whereas \ce{CaO} is identified as a compound and solid \ce{O2} is classified as miscellaneous and excluded from further analysis.

Structural comparisons were performed both within individual ANX formula types and across broader collections of materials. Restricting comparisons within ANX classes prevents chemically incompatible assignments that can arise from purely geometric matching. For example, if atomic identity is ignored, the occupied sites in the CsCl (B2) structure form the same body-centred cubic arrangement as elemental metals despite representing fundamentally different chemical organizations. Comparisons within common stoichiometric classes therefore preserve chemically meaningful framework assignments while retaining the advantages of species-independent geometric matching.
%Structural comparisons were performed both across and within categories of materials sharing the same ANX formula type. Although the framework matching itself is species-independent, unrestricted comparisons across all stoichiometries can merge geometrically similar but chemically distinct structures. For example, if atomic identity is ignored, the occupied sites in the \ce{CsCl} (B2) structure form the same body-centered cubic arrangement (BCC) as elemental metals, despite representing fundamentally different chemical organizations. Restricting comparisons to common stoichiometric classes preserves the advantages of species-independent geometric matching while preventing such chemically incompatible framework assignments.

For naming purposes, formulae were reallocated according to elemental abundance using the geometric mean of crustal abundances of constituent elements~\cite{haynes2016crc}:

\begin{equation}
A = \left( \prod_{i=1}^{n} x_i \right)^{1/n}
\end{equation}

\noindent where $x_i$ is the abundance of element $i$ and $n$ is the number of atoms in the formula unit. The highest-scoring formula within each prototype class was selected as the final representative formula. For example, what could be termed the NaCl structure framework by convention in our case becomes CaO, based on the relative higher levels of abundance. The abundance data used is provided in the SI.

\begin{table}[b]
\centering
\caption{Unique structural frameworks and recurrent prototypes identified for different subsets of the ground-state materials dataset.}
\label{tab:matching-results}
\begin{tabular}{l c c c}
\hline
\textbf{Dataset} & \textbf{Number of Materials} & \textbf{Frameworks} & \textbf{Prototypes} \\
\hline
Full & 23,160 & 6,820 & 2,382 \\
Metals/Intermetallics and Compounds & 11,827 & 1,940 & 907 \\
Metals/Intermetallics & 6,270 & 535 & 327 \\
Compounds & 5,557 & 1,704 & 696 \\
\hline
\end{tabular}
\end{table}

\section{Results}

Establishing appropriate tolerance parameters is essential for reliable identification of prototypes. Relaxed tolerances merge structurally distinct frameworks, whereas overly strict tolerances artificially separate similar structures. To identify a robust operating regime, a sensitivity analysis was performed on the full set of 23,160 experimentally observed ground-state materials prior to filtering.

\autoref{fig:tolerance-sensitivity}a and \autoref{fig:tolerance-sensitivity}b show the number of identified frameworks as functions of the fractional length tolerance, defined as the fractional difference between corresponding lattice-vector lengths, $|l_{1}-l_{2}|/\max(l_{1},l_{2})$, and the angle tolerance, defined as the absolute difference between corresponding interaxial lattice angles, $|\theta_{1}-\theta_{2}|$, respectively. Increasing either tolerance initially produces a rapid reduction in the number of identified frameworks as minor geometric distortions are progressively merged into common structural classes. Beyond a length tolerance of approximately 0.5 and an angle tolerance of $5^\circ$, the framework count stabilizes near 6,800, indicating that framework classification becomes largely insensitive to further increases in either parameter. Length and angle tolerances of 0.5 and $5^\circ$ were therefore adopted for subsequent analysis.

The site tolerance (\autoref{fig:tolerance-sensitivity}(c)) defines the maximum normalized displacement between corresponding atomic sites during structural matching, $|r_1-r_2|/d$, where $d$ is the average free length per atom. Unlike the length and angle tolerances, the site tolerance exhibits both underfitting and overfitting, with a pronounced intermediate plateau again occurring near 6,800 frameworks. At low values, small atomic displacements fragment otherwise equivalent frameworks into multiple classes, whereas large values spuriously merge distinct frameworks as atomic positions become insufficiently constrained. The plateau therefore identifies a stable geometric regime in which framework classification is largely insensitive to the precise choice of site tolerance. A value of 0.3 was selected from the lower end of this plateau to minimize the risk of merging genuinely distinct frameworks. Accordingly, matching tolerances of 0.5, $5^\circ$, and 0.3 were adopted for the length, angle, and site tolerances, respectively, and used throughout this work.

\begin{figure}[t]
	\centering
	\includegraphics[width=\textwidth]{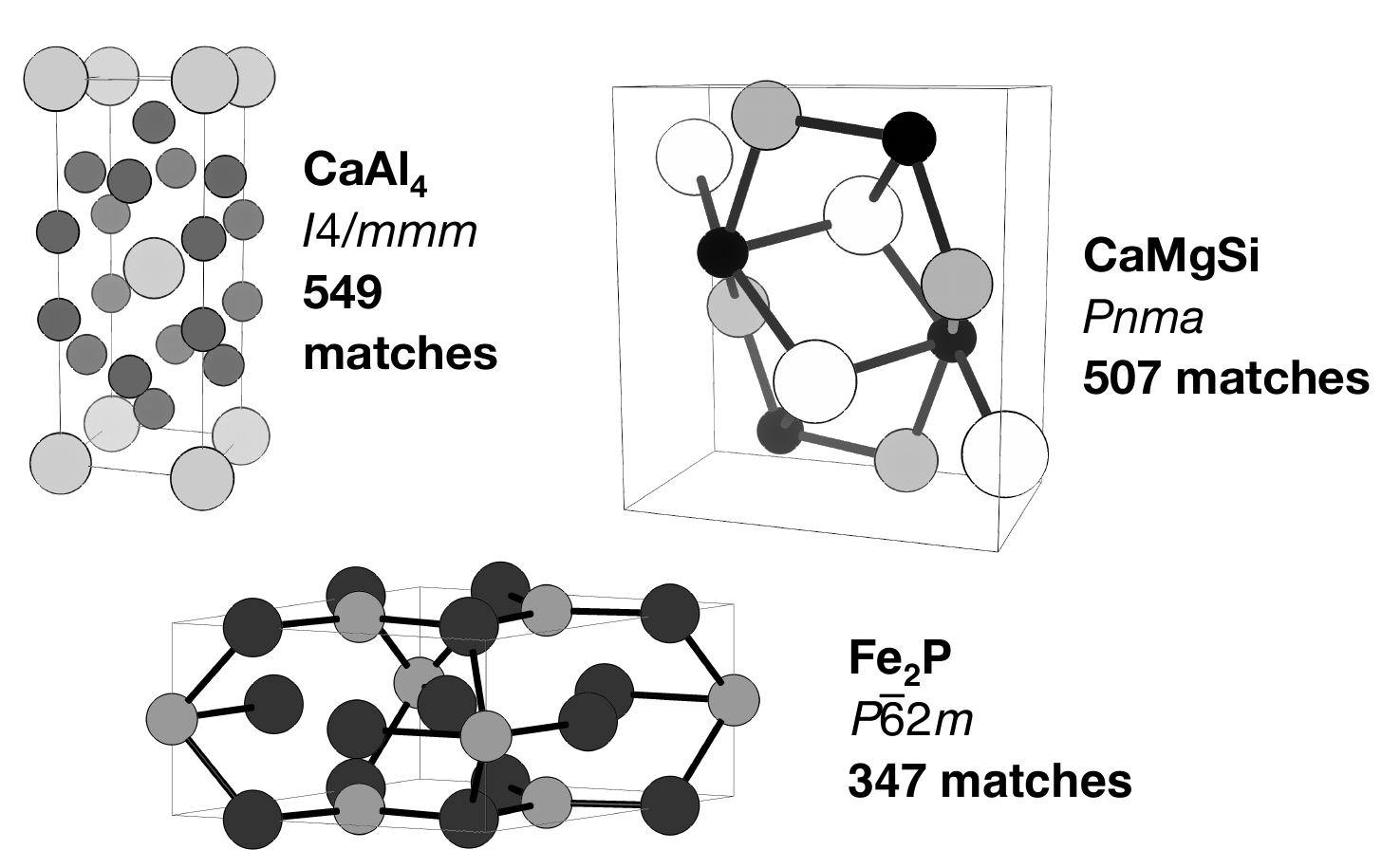}
	\caption{The three most prevalent prototypes among metallic and intermetallic ground-state materials.}
	\label{fig:metal-top-3}
\end{figure}

%Based on the observed stability regions and computational considerations, the final matching tolerances were selected as a length tolerance of 0.5, an angle tolerance of $5^\circ$, and a site tolerance of 0.3. Using these values, the pre- and post-filtered materials were analyzed giving the number of frameworks and results shown in \autoref{tab:matching-results}. 

Application of the filtering criteria reduced the initial dataset of 23,160 Materials Project ground-state structures to 5,557 ionic compounds and 6,270 metallic systems suitable for framework analysis (\autoref{tab:matching-results}). Excluded structures comprised 1,457 hydrogen-containing compounds, 64 noble-gas-containing compounds, 55  miscellaneous systems (see Methods), 496 structures containing more than 100 atoms per unit cell, 9,190 structures with stoichiometric coefficients exceeding four, and 71 structures containing more than three distinct cation or anion species.

\section{Discussion}

The extensive recurrence of structural frameworks observed throughout this study demonstrates that stable ionic materials repeatedly adopt a comparatively small set of preferred structural organizations. Although the number of possible crystal structures is enormous, experimentally realized ground-state materials occupy a highly compressed region of structure space in which a relatively small number of recurring frameworks account for a large and chemically diverse collection of compounds.

%Beginning with the complete set of 23,160 experimentally observed ground-state materials, the analysis identifies 6,820 distinct structural frameworks, of which 2,382 recur across multiple materials and are therefore classified as prototypes. The widespread recurrence of these frameworks demonstrates that stable inorganic materials repeatedly adopt a comparatively small set of preferred arrangements, showing that experimentally realized inorganic structure space occupies a highly compressed subset of the vast crystallographic landscape.

The filtered dataset was subsequently partitioned into metallic/intermetallic systems and ionic compounds to examine how structural recurrence depends on chemical bonding. Although these classes exhibit markedly different bonding characteristics, both remain concentrated within comparatively small sets of recurring structural frameworks. 

%Applying the compositional and structural filters described in the Methods section reduced the dataset to 11,882 materials, within which 1,985 frameworks and 912 prototypes were identified. Although the absolute numbers decrease substantially relative to the full dataset, the overall picture remains unchanged: a large and chemically diverse collection of materials can still be represented by a comparatively limited set of recurring structural frameworks. The filtered dataset was then partitioned into metallic/intermetallic systems and inorganic compounds to examine how structural recurrence varies across these two broad classes of inorganic materials.

\begin{figure}[b!]
	\centering
	\includegraphics[width=\textwidth]{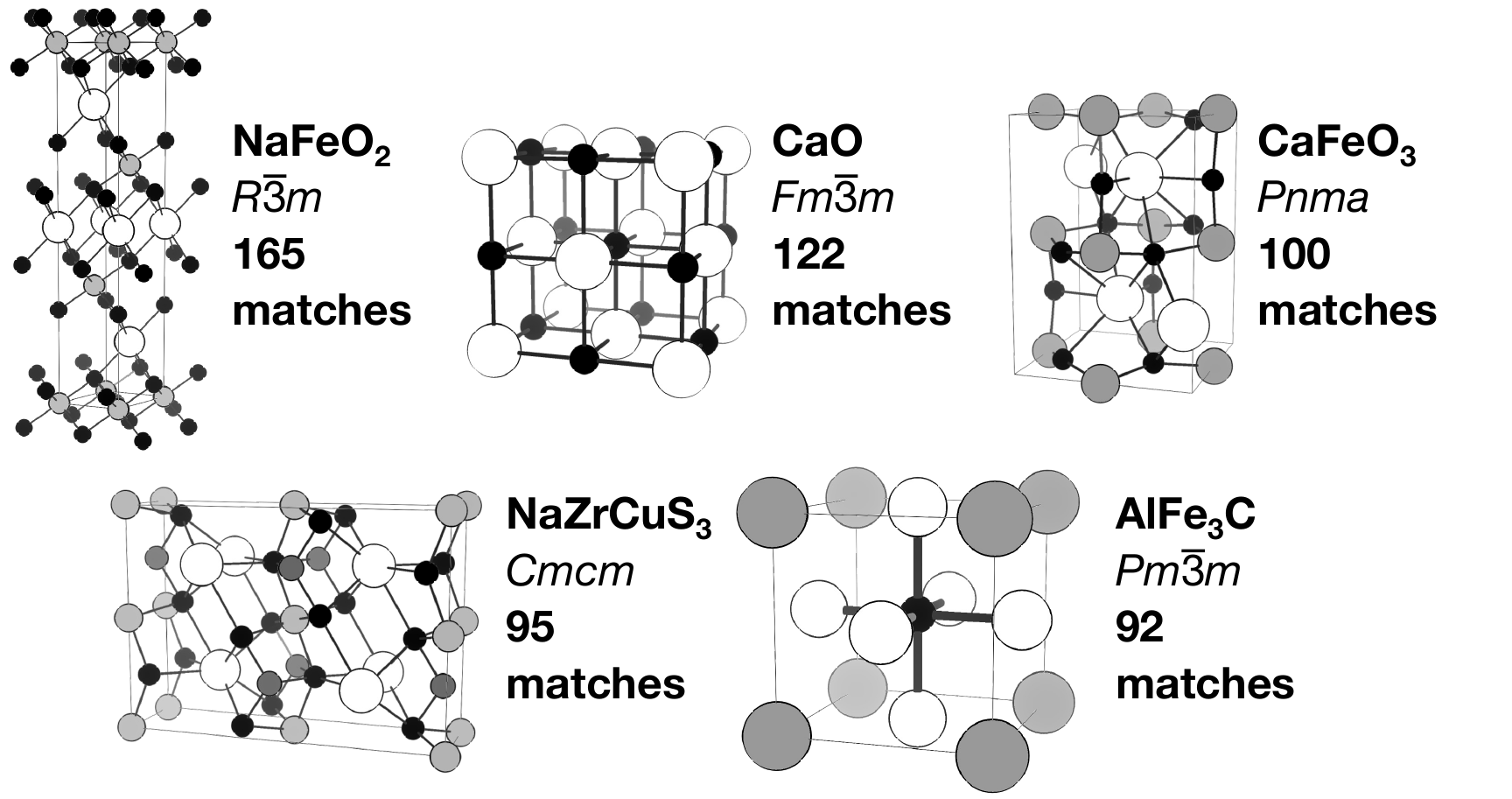}
	\caption{The five most common prototypes among ionic compound ground-state materials.}
	\label{fig:filtered-prototypes}
\end{figure}

The strongest structural recurrence is observed in metallic and intermetallic systems, with 6,270 materials grouped into only 535 frameworks and 327 prototypes. The most populous examples are shown in \autoref{fig:metal-top-3}. This behavior is consistent with the nature of metallic bonding, which is intrinsically non-directional and therefore places fewer geometric constraints on atomic arrangement than ionic or covalent bonding. Rather than being localized between atoms at specific angles, the shared electrons are delocalized throughout the structure. Furthermore, unlike ionic compounds, metallic systems are not constrained by charge-balance requirements, permitting a broader range of compositions to adopt the same underlying framework. As a consequence, many metallic systems adopt variations of a relatively limited set of packing motifs and their derivatives (i.e. face-centered cubic, body-centered cubic, and hexagonal close packing). The high degree of framework recurrence therefore reflects the comparatively limited number of geometrically efficient atomic arrangements compatible with metallic bonding.

The compound subset contains 1,704 unique frameworks among 5,557 materials, of which 696 satisfy the recurrence criterion for prototype classification. The most prevalent prototype families, shown in \autoref{fig:filtered-prototypes}, illustrate that chemically diverse compounds repeatedly realize the same underlying structural organizations. Unlike metallic systems, compounds must simultaneously satisfy stoichiometric, electrostatic, coordination, and bonding constraints, all of which might be expected to increase structural diversity. The persistence of extensive framework recurrence despite these competing constraints suggests that the number of geometrically favorable solutions is far smaller than the vast number of crystallographic arrangements that are theoretically possible.

%Despite the additional constraints imposed by stoichiometry and chemical bonding, inorganic compounds also exhibit substantial framework recurrence. The compound subset contains 1,704 unique frameworks among 5,557 materials, of which 696 satisfy the recurrence criterion for prototype classification. A relatively small number of prototypes account for a substantial fraction of these materials, as illustrated in \autoref{fig:filtered-prototypes}. Unlike metallic systems, compounds must simultaneously satisfy stoichiometric, electrostatic, coordination, and bonding constraints, all of which might be expected to increase structural diversity. Nevertheless, substantial structural recurrence persists, indicating that chemically complex compounds repeatedly converge on a comparatively limited set of framework families. This suggests that the number of geometrically favorable solutions satisfying these competing constraints is far smaller than the vast number of crystallographic arrangements that are theoretically possible.

\begin{figure}[t!]
  \centering
  \includegraphics[width=\linewidth]{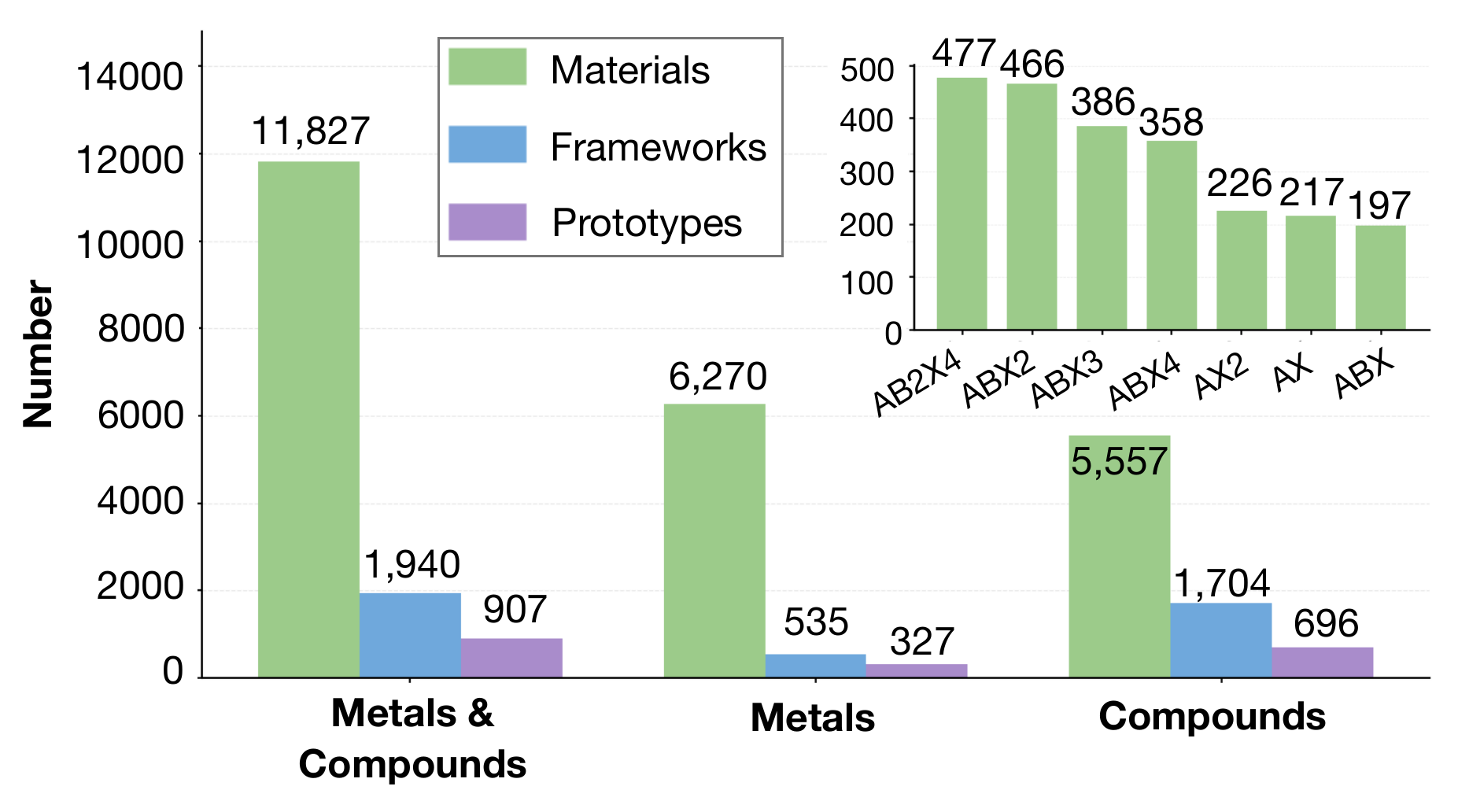}
  \caption{Compression of experimentally realized ground-state materials within framework and prototype space. The main panel compares the numbers of materials, structural frameworks and prototypes for the complete filtered dataset, metallic/intermetallic systems and ionic compounds. The inset shows the seven most common ANX stoichiometry types ranked by the number of constituent materials, illustrating the concentration of materials within a relatively small number of compositional families.}
  \label{fig:most-popular-ANX-group}
\end{figure}

%The concentration of materials within a limited number of stoichiometric classes is mirrored by an even stronger concentration within framework space. 

%The extent of framework recurrence observed in this work is particularly notable given the vast number of crystallographic arrangements that are, in principle, possible. Even highly simplified combinatorial estimates based on Wyckoff-position occupancy yield numbers that exceed $10^8$ candidate structural arrangements. Although such estimates greatly over-count physically meaningful structures and do not account for geometric equivalence, chemical plausibility, or independent positional degrees of freedom, they nevertheless illustrate the enormous disparity between the vast number of crystallographic arrangements that are theoretically accessible and the comparatively small set of prototypes realized among experimentally observed ground-state materials.

The extent of framework recurrence observed in this work is particularly notable given the vast number of crystallographic arrangements that are, in principle, possible. Even highly simplified combinatorial estimates based on Wyckoff-position occupancy yield numbers exceeding $10^8$ candidate structural arrangements. Although such estimates greatly over-count physically meaningful structures and do not account for geometric equivalence, chemical plausibility, or independent positional degrees of freedom, they nevertheless illustrate the enormous disparity between the vast number of crystallographic arrangements that are theoretically accessible and the comparatively small set of prototypes realized among experimentally observed ground-state materials. For comparison, AFLOW XtalFinder identifies 15,205 frameworks among 34,820 unique materials that came from 60,390 ICSD entries, of which 2,050 are included in the AFLOW Encyclopedia of Crystallographic Prototypes~\cite{hicks2021xtalfinder,eckert_aflow_2024}. Although direct comparison is complicated by differences in both scope and methodology, as AFLOW includes unary through septenary systems and structures beyond experimentally observed ground-state materials, both studies indicate that large collections of materials can be represented by substantially smaller sets of structural frameworks. By restricting the analysis to experimentally realized ground-state materials and requiring recurrence across chemically distinct systems, the present work extends this concept by not only classifying structural frameworks but also quantifying their prevalence, enabling prototype families to be both categorized and prioritized.

\begin{table*}[t]
\centering
\caption{Most common space groups within the metallic/intermetallic (left) and ionic compound (right) datasets. The top ten space groups are shown together with their space-group numbers, occurrence counts, and percentages of the corresponding dataset.}
\begin{subtable}[t]{0.48\textwidth}
\centering
\phantomcaption
\begin{tabular}{lccc}
\hline
Space Group & SG Number & Count & \% \\
\hline
\textit{Pnma} & 62 & 885 & 14.1 \\
\textit{I4/mmm} & 139 & 619 & 9.9 \\
\textit{P6$_3$/mmc} & 194 & 456 & 7.3 \\
\textit{Cmcm} & 63 & 408 & 6.5 \\
\textit{Pm$\bar{3}$m} & 221 & 374 & 6.0 \\
\textit{P$\bar{6}$2m} & 189 & 346 & 5.5 \\
\textit{Fm$\bar{3}$m} & 225 & 267 & 4.3 \\
\textit{P4/nmm} & 129 & 241 & 3.8 \\
\textit{F$\bar{4}$3m} & 216 & 209 & 3.3 \\
\textit{P6/mmm} & 191 & 171 & 2.7 \\
\hline
\textbf{Top 10 total} & -- & \textbf{3,976} & \textbf{63.4} \\
\hline
\end{tabular}
\end{subtable}
\hfill
\begin{subtable}[t]{0.48\textwidth}
\centering
\phantomcaption
\begin{tabular}{lccc}
\hline
Space Group & SG Number & Count & \% \\
\hline
\textit{Pnma} & 62 & 789 & 14.2 \\
\textit{P2$_1$/c} & 14 & 503 & 9.1 \\
\textit{R$\bar{3}$m} & 166 & 287 & 5.2 \\
\textit{Cmcm} & 63 & 252 & 4.5 \\
\textit{C2/m} & 12 & 208 & 3.7 \\
\textit{C2/c} & 15 & 208 & 3.7 \\
\textit{P4/mmm} & 123 & 184 & 3.3 \\
\textit{I4/mmm} & 139 & 178 & 3.2 \\
\textit{P$\bar{1}$} & 2 & 167 & 3.0 \\
\textit{P6$_3$/mmc} & 194 & 162 & 2.9 \\
\hline
\textbf{Top 10 total} & -- & \textbf{2,938} & \textbf{52.9} \\
\hline
\end{tabular}
\end{subtable}
\label{tab:space-group-comparison}
\end{table*}

%For comparison, AFLOW XtalFinder identifies 15,205 frameworks among 34,820 unique materials and 60,390 ICSD entries, of which 2,050 are included in the AFLOW Encyclopedia of Crystallographic Prototypes~\cite{hicks2021xtalfinder,eckert_aflow_2024}. Direct comparison is complicated by substantial differences in both scope and methodology, as AFLOW includes unary through septenary systems and structures beyond experimentally observed ground-state materials. By restricting the analysis to experimentally realized ground-state materials and requiring recurrence across distinct systems, the present work not only provides a framework classification system but also quantifies the prevalence of each prototype, enabling structural frameworks to be both categorized and prioritized.

The ICSD reports approximately 9,015 structure types among 159,000 crystallographic entries~\cite{zagorac_recent_2019}. As with AFLOW, direct comparison is complicated by differences in both scope and methodology. The ICSD includes metastable phases, hybrid organic/inorganic materials, and multiple experimental determinations of the same material system. Furthermore, structure-type assignment in the ICSD relies primarily on crystallographic metadata, including space-group symmetry, Wyckoff-position sequences, and lattice-parameter relationships as discussed in the Introduction~\cite{allmann2007introduction}. By contrast, the present work identifies structural frameworks directly from atomic geometries using species-independent structural matching within common compositional classes. Consequently, recurrence reflects repeated realization of a framework across chemically distinct compounds rather than repeated reporting of related crystallographic entries, providing a measure of structural prevalence rather than database occurrence.

Despite their methodological differences, AFLOW, the ICSD, and the present work all support the broader conclusion that a relatively small number of structural prototypes account for a disproportionately large fraction of experimentally realized materials.  Consequently, continued exploration of composition space is likely to produce many new compounds without requiring a proportional expansion in the number of underlying structural frameworks, further motivating the identification and prioritization of recurring structural prototypes rather than their blind enumeration.

Analysis of the space-group distributions for the metallic/intermetallic and ionic compound datasets (\autoref{tab:space-group-comparison}) reveals both commonality and divergence between the two material classes. In both datasets, the orthorhombic \textit{Pnma} space group is by far the most prevalent, accounting for approximately 14 \% of all materials despite its intermediate symmetry. Beyond this shared preference, however, the distributions diverge substantially. Metallic/intermetallic systems are dominated by comparatively high-symmetry space groups associated with efficient packing motifs, whereas ionic compounds exhibit a broader distribution that includes numerous monoclinic and triclinic groups. Nevertheless, the ten most common space groups account for more than half of the materials in both datasets, indicating that experimentally realized ionic materials remain concentrated within a relatively limited subset of crystallographic symmetries despite their chemical diversity. These contrasting symmetry distributions support the separate analysis of metallic/intermetallic systems and ionic compounds, indicating that the two different bonding regimes underpin distinct regions of crystallographic space despite both exhibiting extensive framework recurrence.

Comparison with widely used prototype collections (\autoref{tab:prototype-comparison}) shows that the vast majority of commonly recognized inorganic structure types are recovered by the present methodology. The few apparent omissions reflect deliberate methodological choices rather than limitations of the structural matching procedure. 
For example, Garnet is absent because its formula type (\ce{A2B3C3X12}) exceeds the compositional complexity limits imposed during dataset filtering rather than because of any limitation in the structural matching algorithm. %Diamond is identified in the unrestricted framework analysis but merges with the geometrically equivalent zinc-blende framework in the metallic subset, where elemental structures are not partitioned by stoichiometric class. %The tetragonal \ce{BaTiO3} polymorph is merged with the cubic perovskite framework under the adopted matching tolerances, reflecting the intended robustness of the classification to small structural distortions.  
The close correspondence with widely recognized prototype collections provides further confidence that the recurrence-based methodology captures the principal structural families underpinning inorganic chemistry. Rather than replacing established prototype nomenclature, the present approach recovers these familiar structures through an automated and reproducible procedure while simultaneously identifying the methodological reasons for the small number of apparent discrepancies.

\begin{table}[b]
\centering
\caption{Comparison of the recurrence-based prototype catalog with widely used curated prototype collections, together with prototype prevalence and compatible ANX formula types.}
\label{tab:prototype-comparison}
\begin{tabular}{ccccc}
\hline
\textbf{Conventional Name} & \textbf{In This Work?} & \textbf{Proposed Name} & \textbf{Count} & \textbf{Formula Type(s)} \\
\hline
\textit{Rock-Salt} & True & \ce{CaO} & 122 & AX \\
\textit{\ce{CaFe2O4}} & True & \ce{NaTi2O4} & 79 & \ce{AB2X4} \\
\textit{Spinel} & True & \ce{Al2FeO4} & 73 & \ce{AB2X4}, \ce{A3X4} \\
\textit{Perovskite} & True & \ce{SrFeO3} & 32 & \ce{ABX3}, \ce{AB3X} \\
\textit{HCP} & True & \ce{Na} & 26 & A \\
\textit{Rutile} & True & \ce{SnO2} & 25 & \ce{AX2}, \ce{ABX}, \ce{ABX4}, \ce{A2X} \\
\textit{FCC} & True & \ce{Al} & 16 & A \\
\textit{Diamond} & True & \ce{Si} & 13 & A, AX \\
\textit{BCC} & True & \ce{Fe} & 7 & A \\
\textit{\ce{CaB2O4}} & True & \ce{SrB2O4} & 2 & \ce{AB2X4} \\
%\textit{Tetragonal \ce{BaTiO3}} & False & - & - & \ce{ABX3} \\
\hline
\end{tabular}
\end{table}

As expected for a recurrence-based classification, many structural frameworks are realized across multiple compatible stoichiometries. Our prototype catalogue therefore records framework occurrence at multiple levels of chemical specificity: framework identification is performed independently within individual ANX classes to prevent chemically incompatible assignments, while equivalent framework families identified within separate ANX classes are also linked to reveal their recurrence across inorganic chemistry. Unlike traditional prototype collections, this hierarchical representation shows how structural frameworks span multiple stoichiometric classes while preserving chemically meaningful classifications. Rutile, for example, occurs within \ce{AX2}, \ce{ABX}, \ce{ABX4}, and \ce{A2X} systems, while spinel appears in both the classical \ce{AB2X4} stoichiometry and \ce{A3X4} compounds such as magnetite. Such relationships highlight the geometric generality of recurring frameworks beyond any single representative composition.

A key advantage of the recurrence-based approach is that each prototype is naturally associated with an occurrence frequency. Unlike conventional curated prototype collections, which implicitly treat all structure types as equally representative, the present analysis quantifies how frequently each framework is realized among experimentally realized ground-state materials within its corresponding compositional class. This provides a physically motivated ranking of structural frameworks, allowing candidate templates to be prioritized for element substitution, crystal structure prediction, and high-throughput computational screening according to their empirical prevalence. The resulting recurrence-based prototype catalog and consistent approach to material categorization also provide a reproducible benchmark for evaluating crystal structure similarity metrics, structural descriptors, and machine-learning representations of inorganic solids.

\section{Conclusion}

This work introduces a recurrence-based approach to structural prototype classification in which prototype status is determined by repeated realization among experimentally realized ground-state materials rather than by geometric uniqueness alone. By distinguishing structural recurrence from structural novelty, the resulting framework provides a quantitative description of structural prevalence and demonstrates that experimentally realized inorganic materials occupy a highly compressed region of structure space organized around a comparatively small number of recurring structural frameworks.

Applied to a dataset of 23,160 experimentally observed ground-state inorganic materials, the analysis reveals extensive structural recurrence across all investigated classes of materials. The full dataset contains 6,820 unique frameworks and 2,382 recurrent prototypes, demonstrating that experimentally realized materials occupy only a limited subset of the vast crystallographic landscape that is theoretically accessible. Substantial recurrence is observed in both metallic/intermetallic systems and ionic compounds, indicating that the repeated realization of common structural motifs is a general feature of stable inorganic chemistry rather than a property of a particular class of materials.  

The analysis also highlights important differences between material classes. Metallic and intermetallic systems exhibit particularly strong framework reuse, consistent with the comparatively non-directional nature of metallic bonding and the prevalence of efficient packing motifs. At the same time, ionic compounds remain highly recurrent despite the additional constraints imposed by stoichiometry, coordination environments, electrostatics, and bond directionality. Together, these observations suggest that stable inorganic materials repeatedly adopt a restricted set of geometrically and energetically favorable framework organizations across a broad range of chemical compositions.

The study further highlights challenges associated with classifying low-symmetry and distorted structures, where small geometric variations can obscure underlying framework relationships. Direct geometric comparison therefore provides a useful complement to symmetry- and metadata-based classification schemes by identifying structural equivalence directly from atomic geometries, although at increased computational cost.

The resulting catalog of structural prototypes provides a compact and experimentally grounded description of inorganic structure space. In addition to recovering the canonical prototype families recognized in traditional curated collections, the catalog quantifies their prevalence among stable inorganic materials and identifies the compatible ANX formula types over which they recur. By combining geometric recurrence with empirical occurrence frequencies, the catalog provides both a structural classification scheme and a physically motivated ranking of prototype importance for materials discovery, crystal structure prediction, and materials informatics.

Overall, this work demonstrates that experimentally realized inorganic materials occupy a remarkably constrained region of structure space. Although the number of chemically accessible compounds is vast and continues to grow with increasing compositional complexity, the underlying structural diversity is organized around a comparatively small number of recurring geometric frameworks whose prevalence can now be quantified systematically. Continued exploration of composition space is therefore likely to expand the diversity of known compounds far more rapidly than the diversity of underlying structural frameworks, making recurrence-based prototype classification an increasingly valuable means of categorizing and prioritizing candidate structures for future materials discovery.

\section*{Acknowledgements}

The authors would like to thank Profs. Semën Gorfman and Oswaldo Dieguez for useful discussion. LAB acknowledges support from the Israeli Science Foundation, grant numbers 1415/25 and 1416/25. 

\section*{Supporting Information}

The Supporting Information contains the complete catalog of structure prototypes identified in this study, together with their associated structural metadata. The dataset is also available at \url{https://github.com/bmd-lab/structure-prototypes}.

%%%%%%%%%%%%%%%%%%%%%%%%%%%%%%%%%%%%%%%%%%%%%%%%%%%%%%%%%%%%%%%%%%%%%
%% If you are using classical BibTeX rather than biblatex,
%% remove the \printbibliography and uncomment the \bibliograpy one
%%%%%%%%%%%%%%%%%%%%%%%%%%%%%%%%%%%%%%%%%%%%%%%%%%%%%%%%%%%%%%%%%%%%%
\printbibliography
%\bibliography{references.bib}

% \newpage

% \rule{0.05in}{1.75in}%
% \begin{minipage}[b][1.75in]{3.25in}
%   \sffamily
%   \frenchspacing

%   Some journals require a graphical entry for the Table of Contents. This
%   should be laid out ``print ready'' so that the sizing of the text is correct.

%   The space available depends on the journal: J. Am. Chem. Soc. allows 3.25 in
%   by 1.75 in and requires sanserif text. Some journals want different sizes:
%   you can easily adjust here.
  
%   The two rules either side of the content are there to help judge the height
%   of your material: they may be deleted once not required.
  
% \end{minipage}%
% \rule{0.05in}{1.75in}

\end{document}